# The magic of disc-worlds: non-rotating methanol masers

**Huib Jan van Langevelde[*], Karl J.E. Torstensson**
*Joint Institute for VLBI in Europe, Postbus 2, 7990 AA Dwingeloo the Netherlands*
*Sterrewacht Leiden, Leiden University, Postbus 9513, 2300 RA Leiden, the Netherlands*
*E-mail:* `langevelde@jive.nl, kalle@strw.leidenuniv.nl`

**Anna Bartkiewicz, Marian Szymczak**
Torun Centre for Astronomy, Nicolaus Copernicus University, Poland
*E-mail:* `annan@astro.uni.torun.pl, msz@astro.uni.torun.pl`

**Wouter H.T. Vlemmings, Gabriele Surcis**
*Argelander-Institut für Astronomie der Universität Bonn, Auf dem Hügel 71, 53121 Bonn, Germany*
*E-mail:* `wouter@astro.uni-bonn.de, gsurcis@astro.uni-bonn.de`

**Andreas Brunthaler**
*Max-Planck-Institut für Radioastronomie, Auf dem Hügel 69, 53121 Bonn, Germany*
*E-mail:* `brunthal@mpifr-bonn.mpg.de`

In recent studies of methanol masers, a substantial fraction of the objects show maser components aligned in large-scale elliptical configurations. These can be readily interpreted as rings centred on a high mass star in formation, seen in projection. Remarkably, most of these rings do not show signs of rotation, but rather the radial motions dominate. This must mean that their dynamics are governed by other than gravitational forces. In particular, we have studied the methanol masers around Cep A in detail, where it can be argued that the methanol masers show signs of infall. In this paper we discuss the dynamics of the Cep A methanol maser and sources from the Torun blind survey to argue that at least in a fraction of sources methanol masers could be associated with the shock interface between the large scale accretion, regulated by the magnetic field, and a 1000-AU scale circumstellar disk. We discuss the validity of such a model for the overall population of methanol maser sources.

[*] Speaker

## 1. Introduction

From their first discovery it has been clear that methanol masers are uniquely associated with high mass star formation [9]. For a significant fraction of sources an association with an early HII region has been found [21] and in the vicinity of most other methanol masers one finds high mass protostellar clumps [7]. Because of their association with high mass stars and the possibility to obtain high resolution kinematic information, not hindered by dust extinction, methanol masers are potentially powerful probes of the detailed process of high-mass star formation. As the evidence is building up that high mass stars form in a similar way as their, much more directly observable, low mass counterparts, it is particularly interesting that methanol masers can even be used to map out the magnetic field in these objects. However, the complexity of high-mass star formation has so far prevented a clear association of this maser phenomenon with a specific physical component of the young star. There have been claimed associations with disc components, as well as outflows [12][13][11]. Understanding the origin of the methanol masers in these objects is also important for constraining the excitation mechanism. Clearly the methanol masers require a substantial methanol abundance, probably obtained by heating or destroying interstellar grains on which methanol can form. In addition a high temperature and enhanced density are deemed necessary, together with an IR field to excite the maser [5]. Understanding the precise nature of methanol masers is also important for their use as probes of the Galactic structure and dynamics, because the internal velocities should ideally be separated from the Galactic rotation motions [14].

## 2. Observing Rings and Ellipses

In a previous study [3], we have imaged methanol masers, selected from the blind Torun survey [15] with the EVN[1]. The EVN has unique high resolution imaging capabilities at 6.7 GHz and for most of these observing campaigns an 8 or 9 antenna array was available. Moreover, after initial observing with MERLIN, it was possible to do these observations in astrometric mode, allowing us to make detailed associations with diagnostics at other wavelengths and in particular other maser species [2].

The remarkable result from these observations was that no less than 33% of the masers showed maser clumps distributed in ellipses, suggesting a ring shaped morphology projected at various inclinations. Other morphologies were simple (single, double, linear), but also more complex morphologies were detected that cannot be consistent with such ring structures. In the past several authors have interpreted linear masers as being part of a circumstellar disk seen edge-on, arguing that the observed velocity gradients are due to Keplerian rotation (e.g. [13]).

However, the elliptical structures observed with the EVN in our study do not display the expected rotation signature in which the highest velocities occur on the extreme points on the long axis of the ellipses. Recently, we found a similar result in the archetypical high mass protostellar object Cep A HW2 [18]. In this source the distribution of the masers delineates a 1730-mas structure, straddling the waist of the bipolar HII region, suggesting all the masers originate in a large scale ring structure perpendicular to the ionized outflow (Figure 1). The positions of the masers can be convincingly fit to a ring, with the centre projected very close to the position of the central radio-continuum source [6]. Although some of the individual maser clusters have large velocity gradients, the large-scale velocity structure is inconsistent with a rotating disk. In fact, fitting the velocities gives a better result when we allow for radial motions, instead of rotation.

---

[1] The European VLBI Network is a joint facility of European, Chinese, South African and other radio astronomy institutes funded by their national research councils.

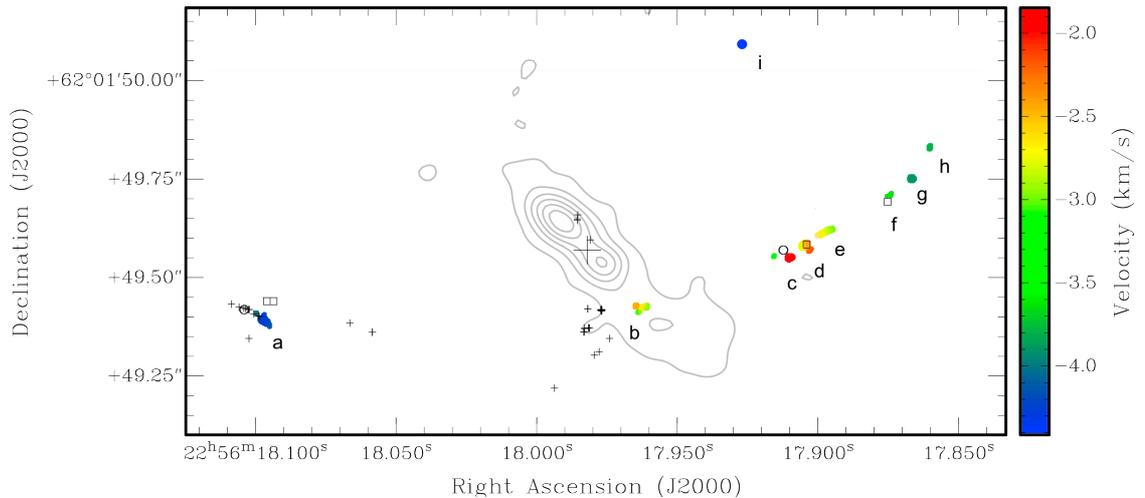

*Figure 1 The Cep A VLA continuum [17], with the methanol masers plotted with colours indicating the line of sight velocities.*

A similar result is found for the elliptical sources in the larger sample. For these, the radial motion can point to either infall or expansion, depending on the sign of the inclination. At least for Cep A there are two arguments that the inclination is infall. First, the outflow in Cep A is such that the north side is known to be the approaching side; the ring of masers could then be perpendicular to this if the southern part of the ring is nearest to the observer. Secondly, the southern side is also the brightest part of the maser ring, with fewer and less bright masers observed from the northern side. Absorption by the ionized medium in which the jet occurs could be responsible for the blockage of the backside masers, although the HII region does not appear to be optically thick at 6.7 GHz. Another explanation could be that the front side masers are brighter because they amplify the background radio continuum.

For Cep A and one source from the EVN sample (G23.657-0.127)[1] we have additional information from 12 GHz monitoring with the VLBA[2] [18][4]. This allows us to search for maser proper motions to get an independent handle on the internal kinematics of these sources. For Cep A the proper motions are consistent with no tangential motion. For the ring source G23.657-0.127 the situation is slightly different as it is seen almost face on. However, again the proper motions (Figure 2) do not allow for the detection of any systematic proper motion, neither tangential, nor radial, although significant proper motions are detected with rather random orientations.

---

[2] The VLBA is operated by NRAO. The National Radio Astronomy Observatory is a facility of the National Science Foundation operated under cooperative agreement by Associated Universities, Inc.

The 12 GHz monitoring of both sources has also yielded parallaxes and distances [10][18][4]. In both sources a central source has been detected and thus we can estimate the luminosity and compare the physical properties of the objects directly. The ring of Cep A (at 700 pc) appears to be 1210 AU across, while for G23.657-0.127 (at 3.19 kpc) we measure 812 AU, both rather large structures. Based on bolometric luminosity estimates, both ring structures have comparable central objects of 18 and 20 $M_\odot$. At the ring location one therefore expects rotation velocities of 5.1 – 6.5 km/s, which we have not been able to infer from the radial motions or the proper motions. Clearly forces other than gravity govern the dynamics of the gas.

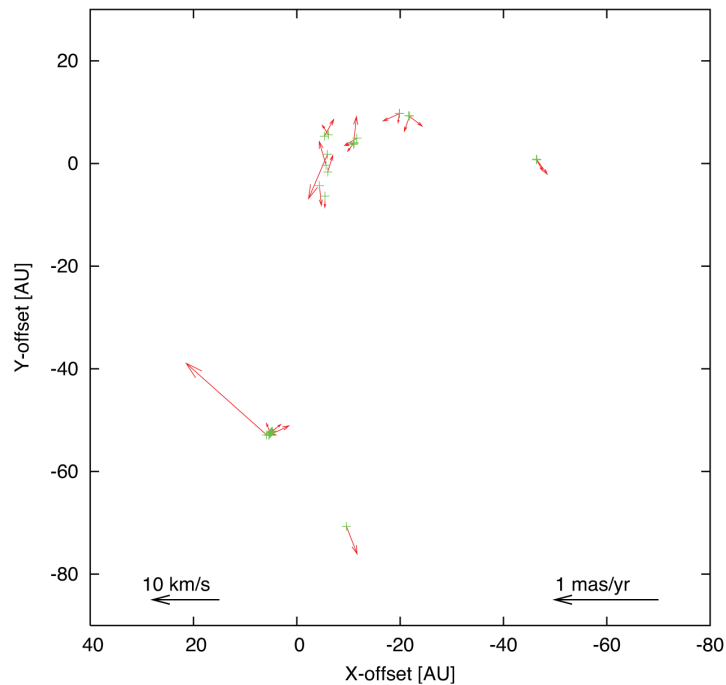

*Figure 2. Residual proper motions of 12 GHz masers observed with the VLBA in the ring source G23.657-0.127, after subtraction of the common Galactic motion and parallax.*

Recently, we have been able to measure the magnetic field strength and direction in the Cep A methanol maser, determining that it could be the dominant force in the ring [19][20]. Moreover, the field direction seems to suggest that magnetic fields are regulating the large-scale accretion onto the circumstellar masers.

## 3. Discussion: An accretion model for methanol masers

For Cep A, and implicitly for the other ring sources, a picture is emerging in which the methanol masers originate on the interface between the large-scale accretion flow and a circumstellar disk, detected in a number of molecular tracers [8]. The infall appears to be governed by the magnetic field. Apparently, the methanol gas is located in the pre-shock gas, where the dynamics are largely governed by the accretion flow. Interestingly, we note that several maser clusters have large velocity gradients and we wonder whether these are the signatures of the shocks interacting with the rotating gas in the disk. Cep A is also the site of abundant water masers [16]; most of these seem to be related to a molecular outflow with a large opening angle, confirming the proposed geometry. In one location the methanol and water masers co-exist, but with a large velocity difference, which could be explained if the water masers reside in the post-shocked gas.

The proposed model is based on magnetic field measurements in a few sources and radial motions in a small sample of ring sources. It is therefore not obvious that the same model holds for all methanol maser sources. We note that single component, double and linear masers in the sample could in principle originate from incomplete rings, but clearly at least some of the complex sources require more elaborate models, possibly involving outflows.

However, the model offers some attractive features. For example it naturally explain the fact that methanol masers can be found over rather large (1000 AU) structures, but yet show very modest velocity spreads compared to for example water masers. Also, these large scale accretion structures will be present in high mass proto-stars before the ionisation of HII regions starts, and possibly even before outflow phenomena occur, linking them with the earliest stages of massive star formation. Moreover, the large scale accretion flows we discover, may be not be present at all in low mass stars, naturally explaining the absence of methanol masers in young stars of a few solar mass and below. The shocks could serve as the place where methanol is released to the gas phase and it would therefore be expected that the maser locations are the origin of the methanol production in these sources.

If the methanol masers were associated with the large-scale accretion, one would expect only modest peculiar velocities and proper motions for them, making them even more suitable for studying the large-scale dynamics of the Galaxy. Future observations will be able to test this model in more detail. For example high sensitivity observations, VLBI complemented with eMERLIN or EVLA, will be needed to test whether ring sources are ubiquitous. Where they are found, it is important to estimate size and direction of the magnetic field and also the axis of associated molecular or ionized outflows. Finally, with the expected progress of millimetre interferometers, it should be possible to study the excitation of the methanol in detail at the location of the maser.

## References

[1] Bartkiewicz, A., Szymczak, M. & van Langevelde, H.J., 2005, *Ring shaped 6.7 GHz methanol maser emission around a young high-mass star*. A&A 442 L61

[2] Bartkiewicz, A., Szymczak M., Van Langevelde H.J., De Buizer J.M., Pihlstrom Y., 2010, *Studies of methanol maser rings*, PoS(10th EVN Symposium)003

[3] Bartkiewicz, A. et al., 2009, *The diversity of methanol maser morphologies from VLBI observations*. A&A 502 155

[4] Bartkiewicz, A. et al., 2008, *The nature of the methanol maser ring G23.657-00.127. I. The distance through trigonometric parallax measurements*. A&A 490 787

[5] Cragg, D.M., Sobolev, A.M. & Godfrey, P.D., 2005, *Models of class II methanol masers based on improved molecular data*. MNRAS 360 533

[6] Curiel, S. et al., 2006. *Large Proper Motions in the Jet of the High-Mass YSO Cepheus A HW2*. ApJ 638 878.

[7] Hill, T. et al., 2005. *Millimetre continuum observations of southern massive star formation regions - I. SIMBA observations of cold cores*. MNRAS 363 405

[8] Jiménez-Serra, I. et al., 2009. *Unveiling the Main Heating Sources in the Cepheus a HW2 Region*. ApJ 703 L157

[9] Menten, K.M., 1991. *The discovery of a new, very strong, and widespread interstellar methanol maser line*. ApJ 380 L75

[10] Moscadelli, L. et al., 2009. *Trigonometric Parallaxes of Massive Star-Forming Regions. II. Cep A and NGC 7538*. ApJ 693 406

[11] Moscadelli, L. et al., 2010. *Methanol and water masers in IRAS 20126+4104: The distance, the disk, and the jet*. arXiv1011.4816

[12] Norris, R.P. et al., 1998. *Methanol Masers as Tracers of Circumstellar Disks*. ApJ 508 275

[13] Pestalozzi, M.R. et al., 2004. *A Circumstellar Disk in a High-Mass Star-forming Region*. ApJ 603 L113

[14] Reid, M.J. et al., 2009. *Trigonometric Parallaxes of Massive Star-Forming Regions. VI. Galactic Structure, Fundamental Parameters, and Noncircular Motions*. ApJ 700 137

[15] Szymczak, M. et al., 2002. *6.7 GHz methanol masers at sites of star formation. A blind survey of the Galactic plane between 20° <<= 40° and |b| <= 0°52*. A&A 392 277

[16] Torrelles, J.M. et al., 2010. *A wide-angle outflow with the simultaneous presence of a high-velocity jet in the high-mass Cepheus A HW2 system*. arXiv:1008.2262

[17] Torrelles, J.M. et al., 2007. *The Circumstellar Structure and Excitation Effects around the Massive Protostar Cepheus A HW 2*. ApJ 666 L37

[18] Torstensson, K.J.E. et al., 2010. *Dynamics of the 6.7 and 12.2 GHz methanol masers around Cepheus A HW2*. arXiv1010.4191

[19] Vlemmings, W.H.T., 2008. *A new probe of magnetic fields during high-mass star formation. Zeeman splitting of 6.7 GHz methanol masers*. A&A 484 773

[20] Vlemmings, W.H.T. et al., 2010. *Magnetic field regulated infall on the disc around the massive protostar CepheusAHW2*. MNRAS 404 134

[21] Walsh, A.J. et al., 1998. *Studies of ultra-compact HII regions - II. High-resolution radio continuum and methanol maser survey*. MNRAS 301 640